# Transverse spin and momentum in two-wave interference

Aleksandr Y. Bekshaev[1,2], Konstantin Y. Bliokh[1,3], and Franco Nori[1,4]

[1]*Center for Emergent Matter Science, RIKEN, Wako-shi, Saitama 351-0198, Japan*
[2]*I. I. Mechnikov National University, Dvorianska 2, Odessa, 65082, Ukraine*
[3]*iTHES Research Group, RIKEN, Wako-shi, Saitama 351-0198, Japan*
[4]*Physics Department, University of Michigan, Ann Arbor, Michigan 48109-1040, USA*

We analyze the interference field formed by two electromagnetic plane waves (with the same frequency but different wave vectors), and find that such field reveals a rich and highly non-trivial structure of the local momentum and spin densities. Despite the seemingly-planar and extensively-studied character of the two-wave system, we find that it possesses a *transverse* (out-of-plane) *helicity-independent spin density*, and also a *transverse polarization-dependent momentum density* with unusual physical properties. The polarization-dependent transverse momentum represents the so-called *Belinfante spin momentum*, which does not exert the usual optical pressure and it is considered as 'virtual' in field theory. We perform analytical estimations and exact numerical simulations of the interaction of the two-wave field with probe Mie particles. The results of these calculations clearly indicate the straightforward detectability of the unusual spin and momentum properties in the two-wave field and strongly motivate their future experimental verifications.

## 1. Introduction

It is well-known, since the seminal works by J.H. Poynting [1], that light carries momentum and angular momentum (AM) [2,3]. Typical plane-wave or Gaussian-beam states exhibit longitudinal momentum associated with the wave vector **k** and also longitudinal (**k**-directed) spin AM associated with the degree of circular polarization (helicity) $\sigma$. This is in accordance with the "naïve" but intuitively-clear picture of photons as particles carrying momentum and spin. However, *local* momentum and angular-momentum densities in *structured* (i.e., non-plane-wave) optical fields can demonstrate unusual features, which have recently attracted considerable attention. These are: "super-momentum" with values higher than $\hbar k$ per photon [4–8], transverse (i.e., orthogonal to **k**) helicity-independent spin AM [9–13], and transverse helicity-dependent momentum [10,14,15].

Optical momentum and AM are the main dynamical properties of light, which manifest themselves and play a crucial role in various light-matter interactions [16], including laser cooling [17], optical manipulation of small particles [18], and optomechanical systems [19]. Importantly, momentum and AM of light can be transferred to small absorbing particles or atoms [4,7,8,10,17,18,20–22], generating a radiation-pressure force and torque on the particle [10,23–27]. In other words, the local optical momentum and spin densities can be *measured* via the translational and spinning motion of the probe particles. This was used for the detection of the above-mentioned extraordinary spin and momentum properties in structured fields [4,5,7,8,10–13].

In this work, we are interested in the transverse momentum and spin AM [9–15]. So far, these unusual quantities have been noticed only in *evanescent* waves, i.e., inhomogeneous near-fields defined in half-space and strongly localized in the vicinity of sharp interfaces. An important question is whether the transverse momentum and spin properties can be observed in freely *propagating* far-fields in vacuum: e.g., in usual paraxial laser beams. Here we find that the simplest propagating non-plane-wave field – two interfering plane waves – also exhibits these extraordinary spin and momentum properties. Despite the seemingly planar and thoroughly-studied character of the two-wave system, we discover that such field possesses a *transverse* (i.e., out of the plane formed by the two wave vectors, see Fig. 1) *helicity-independent spin density*, and also a *transverse polarization-dependent momentum* with non-trivial physical properties.

The transverse ($y$-directed in our geometry) quantities are not restricted by the planar $(x,z)$ wave-vector configuration because they are determined by the internal polarization degrees of freedom, which remain truly three-dimensional. Namely, the transverse spin appears due to the phase-shifted (e.g., imaginary) longitudinal $z$-component of the field, while the transverse momentum is also related to the in-plane $x$-inhomogeneity of the field intensity. In the case of evanescent waves [10], the imaginary component of the *single complex wave vector* $\mathbf{k} = k_z \bar{\mathbf{z}} + i\kappa \bar{\mathbf{x}}$ provides for both of these features. In contrast, for propagating waves, considered here, at least *two real wave vectors* $\mathbf{k}_{1,2} = k_z \bar{\mathbf{z}} \pm k_x \bar{\mathbf{x}}$ are needed to generate the truly three-dimensional field and the in-plane $x$-inhomogeneity from the interference (see Fig. 1). Therefore, unlike the evanescent-wave case, where the extraordinary transverse quantities preserve constant signs in the $x>0$ half-space and only decay exponentially together with the field intensity, their counterparts in the two-wave field oscillate and change signs across the interference pattern.

To investigate the manifestations of the transverse momentum and spin AM in light-matter interactions, we calculate the optical forces and torques on a Mie particle immersed in the two-wave interference field. Remarkably, depending on the particle's position and wave polarizations, the particle can experience *transverse torque* about the out-of-plane axis, even in the case of *linearly* in-plane polarized incident waves with zero helicity. Furthermore, the particle can undergo a *transverse force* orthogonal to the wave vectors and strongly dependent on the wave polarization. These intriguing results, supported by both analytical theory and exact numerical simulations, call for experimental verification.

## 2. Momentum and spin in a two-wave interference field

Throughout this paper we consider monochromatic electric and magnetic fields, $\mathcal{E}(\mathbf{r},t) = \mathrm{Re}\left[\mathbf{E}(\mathbf{r})e^{-i\omega t}\right]$ and $\mathcal{H}(\mathbf{r},t) = \mathrm{Re}\left[\mathbf{H}(\mathbf{r})e^{-i\omega t}\right]$, and use Gaussian units. All the properties we discuss below hold in free space, but to conform to optical-manipulation experiments using water or oil, we assume a homogeneous medium with real permittivity $\varepsilon$, permeability $\mu$, and refractive index $n = \sqrt{\varepsilon\mu}$.

The field we consider is a superposition of two plane waves with arbitrary polarizations propagating in the $(x,z)$ plane at an angle $2\gamma$ between their wave vectors (see Fig. 1)

$$\mathbf{k}_{1,2} = k\left(\cos\gamma\,\bar{\mathbf{z}} \pm \sin\gamma\,\bar{\mathbf{x}}\right). \tag{1}$$

Here $k = n\omega/c$, the two signs correspond to the indices 1 and 2, and hereafter $\bar{\mathbf{x}}$, $\bar{\mathbf{y}}$, and $\bar{\mathbf{z}}$ denote the unit vectors of the corresponding axes. The complex electric fields of the two waves can be written as

$$\mathbf{E}_{1,2} = \frac{A}{\sqrt{1+\left|m_{1,2}\right|^2}}\left(\cos\gamma\,\bar{\mathbf{x}} + m_{1,2}\,\bar{\mathbf{y}} \mp \sin\gamma\,\bar{\mathbf{z}}\right)e^{i\Phi_{1,2}}, \tag{2}$$

where $\Phi_{1,2} = k(z\cos\gamma \pm x\sin\gamma)$ are the wave phases and we assume that the two waves have equal real electric-field amplitudes $A$. In equation (2), $m_{1,2}$ are the complex parameters describing the wave polarizations [10,28]. The corresponding normalized Stokes parameters characterizing the degrees of the vertical/horizontal, diagonal $45°/−45°$, and right-hand/left-hand circular polarizations on the Poincaré sphere are, respectively:

$$\tau_{1,2} = \frac{1-|m_{1,2}|^2}{1+|m_{1,2}|^2}, \quad \chi_{1,2} = \frac{2\operatorname{Re}(m_{1,2})}{1+|m_{1,2}|^2}, \quad \sigma_{1,2} = \frac{2\operatorname{Im}(m_{1,2})}{1+|m_{1,2}|^2}. \tag{3}$$

Thus, $\sigma_{1,2}$ are the helicities of the two waves. The wave magnetic fields $\mathbf{H}_{1,2}$ corresponding to (2) are given in Supplemental Material, and the resulting interference fields are $\mathbf{E} = \mathbf{E}_1 + \mathbf{E}_2$ and $\mathbf{H} = \mathbf{H}_1 + \mathbf{H}_2$.

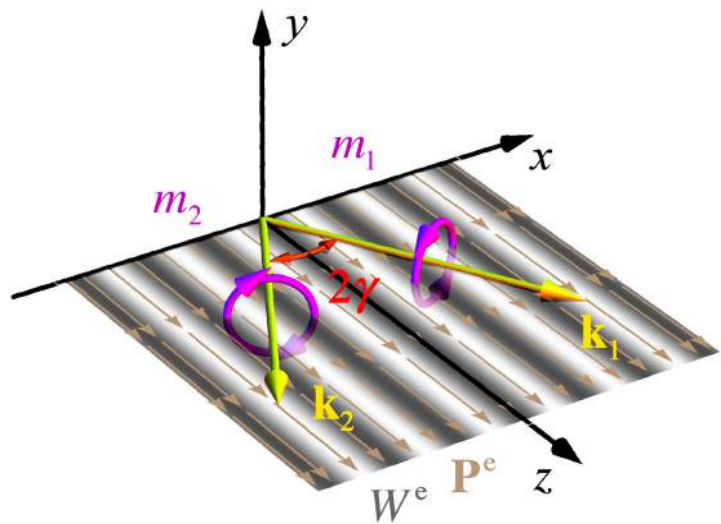

**Figure 1. Interference of two polarized plane waves.** Two waves having equal amplitudes and wave vectors $\mathbf{k}_{1,2}$ with an angle $2\gamma$ between them propagate and interfere in the $(x,z)$ plane. The wave polarizations are characterized by complex parameters $m_{1,2}$, and here the case of opposite circular polarizations $m_1 = -m_2 = i$ (i.e., the Stokes parameters $\sigma_1 = -\sigma_2 = 1$) is shown. Here and in Figs. 2 and 3 below, the grayscale plot represents the distribution of the electric energy density $W^e(x)$, equations (4) and (9). The in-plane brown arrows show the electric part of the canonical momentum density of light, $\mathbf{P}^e(x)$, equations (5) and (10). This canonical momentum determines the energy transport, optical pressure, and it is directed along the $z$-axis independently of the wave polarizations.

The main local dynamical characteristics of an optical field are the time-averaged densities of energy, $W$, momentum, $\mathbf{P}$, and spin AM, $\mathbf{S}$. They are described by the following equations with separate electric and magnetic contributions [5,10,27,29]:

$$W = W^e + W^m = \frac{g\omega}{2}\left(\varepsilon|\mathbf{E}|^2 + \mu|\mathbf{H}|^2\right), \tag{4}$$

$$\mathbf{P} = \mathbf{P}^e + \mathbf{P}^m = \frac{g}{2}\operatorname{Im}\left[\frac{1}{\mu}\mathbf{E}^*\cdot(\nabla)\mathbf{E} + \frac{1}{\varepsilon}\mathbf{H}^*\cdot(\nabla)\mathbf{H}\right], \tag{5}$$

$$\mathbf{S} = \mathbf{S}^e + \mathbf{S}^m = \frac{g}{2}\operatorname{Im}\left(\frac{1}{\mu}\mathbf{E}^*\times\mathbf{E} + \frac{1}{\varepsilon}\mathbf{H}^*\times\mathbf{H}\right), \tag{6}$$

where $g = (8\pi\omega)^{-1}$ in Gaussian units. It should be emphasized that equation (5) determines the so-called *canonical* (or orbital) momentum of light, which can be associated with the local phase gradient of the field [5], and which follows from the Noether theorem and canonical energy-momentum tensor [29]. This momentum is responsible for the energy transfer, radiation pressure, and also appears in quantum weak measurements of the photon momentum (see [5,7–10,27,29–31]). Thus, it is this canonical momentum, but *not* the Poynting vector, that represents the *directly observable* momentum of light. In particular, the momentum (5) is responsible for the "super-momentum" effects $|\mathbf{P}| > W/c$ [4–8], which are impossible with the Poynting vector that never exceeds $W/c$ in absolute value [9].

Nonetheless, below we will also use the *complex* Poynting momentum $\mathbf{\Pi}$ [2], which plays a role in higher-order light-matter interactions [10] (or interactions with complex particles [27]):

$$\mathbf{\Pi} = \frac{gk}{n}\left(\mathbf{E}^* \times \mathbf{H}\right). \tag{7}$$

The real part of the Poynting vector (7) differs from the canonical momentum (5) by the so-called *spin momentum* $\mathcal{P}_S$, which was introduced in 1939 by Belinfante to explain the spin AM of quantum particles within field theory [32–35] (see also [5,9,10,29,30]):

$$\operatorname{Re}(\mathbf{\Pi}) = \mathbf{P} + \mathcal{P}_S, \quad \mathcal{P}_S = \frac{1}{2}\nabla \times \mathbf{S}. \tag{8}$$

Importantly, the divergenceless spin momentum $\mathcal{P}_S$ does not transfer energy, does not exert optical pressure on spherical dipole particles, and is often considered as "virtual", i.e., non-observable. It is this momentum that appears as the enigmatic Fedorov–Imbert momentum in evanescent waves [10,14,15], which is orthogonal to the wave vector and depends on the polarization helicity.

Substituting electric and magnetic fields of the superposition of two waves (2) into the general equations (4)–(6), we calculate the energy, momentum, and spin AM distributions in the two-wave interference. Formulas for the generic polarizations $m_1$ and $m_2$ are given in [Supplemental Material](), while here, for the sake of simplicity, we consider a particular case of the "mirror-symmetric" polarizations, $m_1 = -m_2 \equiv m$, with the Stokes parameters $(\tau_1, \chi_1, \sigma_1) = (\tau_2, -\chi_2, -\sigma_2) \equiv (\tau, \chi, \sigma)$. Omitting the common factor $\varepsilon g A^2$ in all dynamical characteristics, this yields

$$W^{\text{e,m}} = \omega\left[1 + \left(\tau\cos^2\gamma \mp \sin^2\gamma\right)\cos\Phi\right], \tag{9}$$

$$\mathbf{P}^{\text{e,m}} = \frac{\cos\gamma}{cn}W^{\text{e,m}}\,\overline{\mathbf{z}}, \tag{10}$$

$$\mathbf{S}^{\text{e,m}} = \frac{1}{n^2}\left[\sigma\sin\gamma\left(1 \mp \cos\Phi\right)\overline{\mathbf{x}} + \boxed{\frac{\tau \pm 1}{2}\sin 2\gamma \sin\Phi\,\overline{\mathbf{y}}} + \chi\cos\gamma\sin\Phi\,\overline{\mathbf{z}}\right], \tag{11}$$

where the two signs correspond to the indices "e" and "m", while $\Phi = \Phi_1 - \Phi_2 = 2kx\sin\gamma$ is the phase difference that determines the interference pattern (see Fig. 1). In a similar way, we also find the spin momentum and imaginary Poynting vector, equations (7) and (8):

$$\mathcal{P}_S = \frac{k\sin 2\gamma}{n^2}\left[\boxed{-\chi\cos\Phi\,\overline{\mathbf{y}}} + \tau\sin\gamma\cos\Phi\,\overline{\mathbf{z}}\right], \tag{12}$$

$$\mathrm{Im}(\mathbf{\Pi}) = -\frac{2k\sin\gamma}{n^2}\left[\sin\Phi\,\bar{\mathbf{x}} + \boxed{\sigma\cos\gamma\cos\Phi\,\bar{\mathbf{y}}}\right]. \tag{13}$$

Equations (9)–(13) are the key equations of this work. The energy densities and canonical momenta, Eqs. (9) and (10), correspond to the picture of the two-wave interference, which is intuitively clear and known for decades. Namely, $W^{e,m}(x)$ contains the usual interference fringes determined by $\cos\Phi = \cos(2kx\sin\gamma)$, while $\mathbf{P}^{e,m}(x)$ is naturally directed along the $z$ axis and corresponds to the group propagation velocity $v_g = P_z^{e,m}c^2/W^{e,m} = c\cos\gamma/n$ [7], Fig. 1. In contrast, equations (11)–(13) reveal unexpected and counterintuitive dynamical features in such a primitive system. Despite the seemingly planar $(x,z)$ geometry of the problem, the spin AM and the complex Poynting momentum have *transverse out-of-plane $y$-components*.

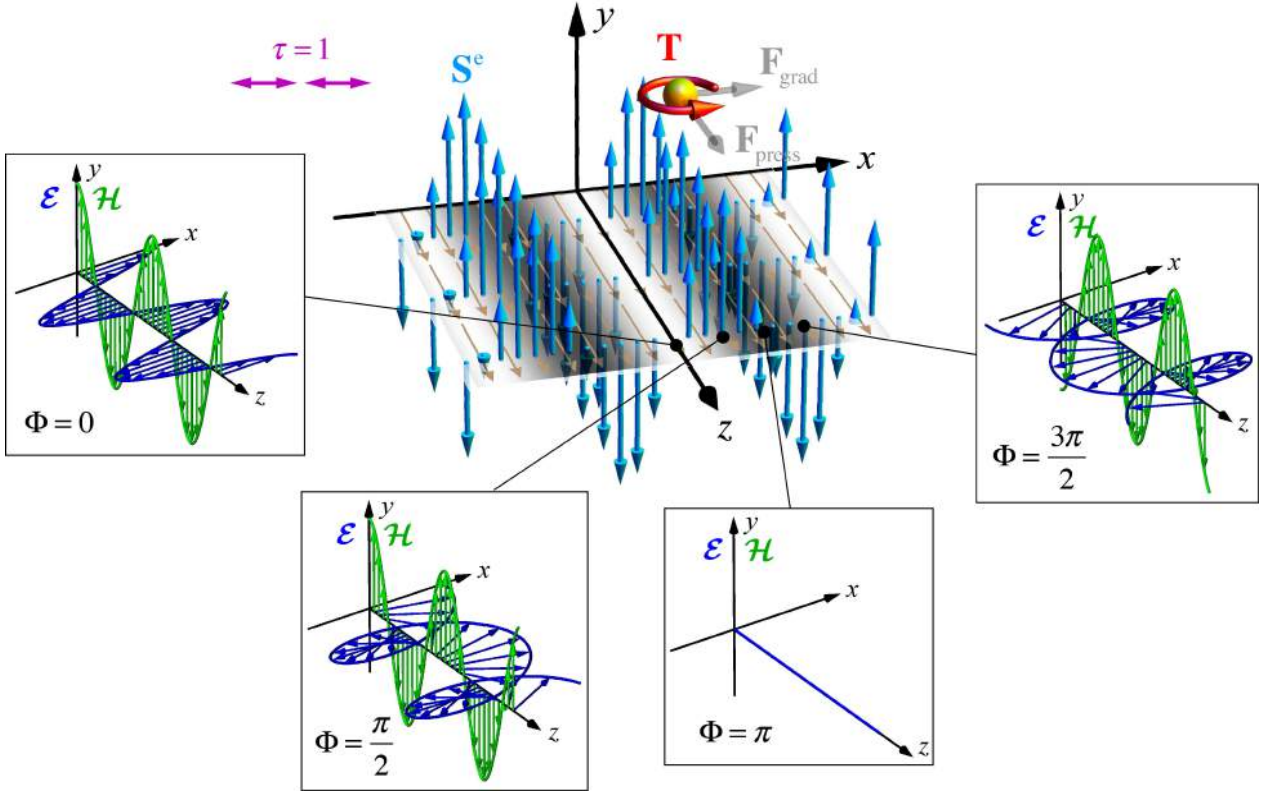

**Figure 2. Transverse helicity-independent spin in the two-wave interference field.** Distribution of the electric spin AM density (11) $\mathbf{S}^e(x)$ is shown here for the simplest case of *linearly* in-plane polarized waves: $m_1 = m_2 = 0$, i.e., the Stokes parameters $(\tau,\chi,\sigma) = (1,0,0)$ [shown schematically in purple]. Despite the seemingly $(x,z)$-planar character of the problem without any helicity, the *transverse $y$-directed spin AM* density appears. The inset panels display instantaneous electric and magnetic field distributions, $\boldsymbol{\mathcal{E}}(\mathbf{r},0)$ and $\boldsymbol{\mathcal{H}}(\mathbf{r},0)$, as functions of $z$ at different $x$-positions, indicated by the values of the phase $\Phi = 2kx\sin\gamma$. These distributions show that the transverse spin arises from the cycloid-like in-plane distribution of the electric field (cf. the evanescent-wave case [9–12]) with the direction of rotation dependent on $\Phi$. An absorbing probe particle is shown here at the position corresponding to $\Phi = \pi/2$, and the optical forces and torques are indicated schematically. The transverse spin AM is locally transferred to the particle, thereby exerting a *radiation torque* (15) $\mathbf{T} \propto \mathbf{S}^e$ [shown in red] (see Fig. 4 for numerical simulations). The particle also experiences in-plane radiation-pressure and gradient forces (14) [shown in grey].

The second term in equation (11) [shown in blue frame] describes the *transverse y-directed spin AM*, which is *independent of the helicity $\sigma$* and can appear even for linear in-plane polarization, Fig. 2. Similar transverse spin was previously described only in evanescent fields [9–13]. This transverse spin density varies sinusoidally across the interference fringes, so that the integral (i.e., $\Phi$-averaged) spin AM vanishes at $\sigma = 0$. The instantaneous $t = 0$ distribution of the electric and magnetic fields $\mathcal{E}(\mathbf{r},t)$ and $\mathcal{H}(\mathbf{r},t)$, shown in Fig. 2, illuminate the origin of this transverse spin $S_y^e$. Interference of the $x$- and $z$-components of the electric wave fields (2) which arrive to the observation point with different phases results in the *cycloidal* field distributions and in-plane *rotation* upon the propagation along the $z$-axis, cf. [9–13] (see also Supplemental Material).

Note that the integral spin AM in the chosen polarization configuration $m_1 = -m_2$ originates from the first term in (11): $\langle \mathbf{S} \rangle = 2\sigma n^{-2} \sin\gamma \, \bar{\mathbf{x}}$. It is proportional to the helicity $\sigma$, as expected, but it is directed along the $x$-axis, i.e., also orthogonally to the main propagation direction. This spin AM is similar to that recently described in [36], and it can be explained by the summation of the usual spin AM from the two waves: $(\sigma_1 \mathbf{k}_1 + \sigma_2 \mathbf{k}_2)/k = 2\sigma \sin\gamma \, \bar{\mathbf{x}}$.

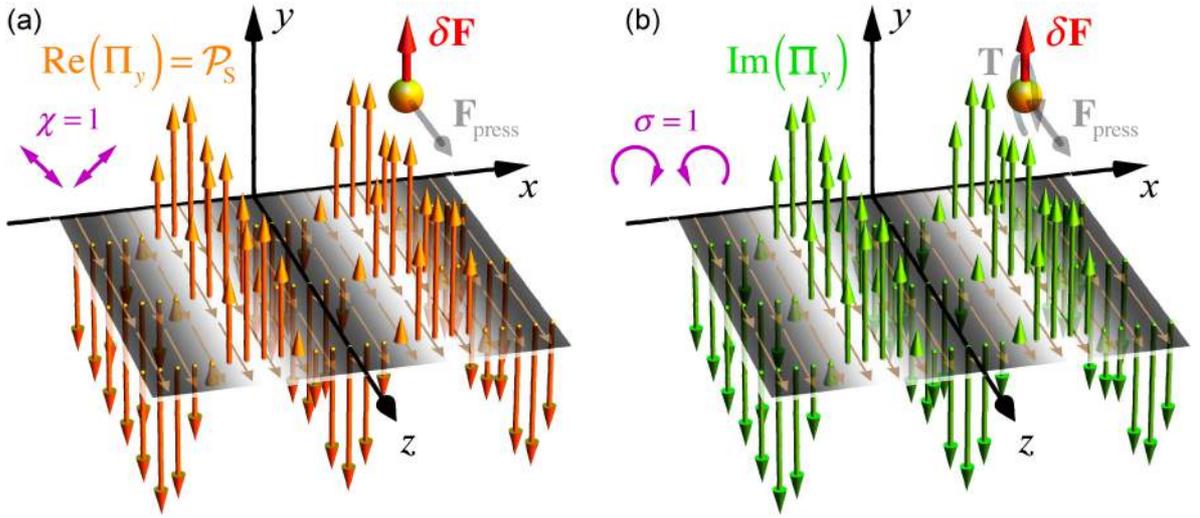

**Figure 3. Transverse polarization-dependent momenta in the two-wave interference field. (a)** Distribution of Belinfante's spin momentum density (12) $\mathcal{P}_S(x)$ ("virtual" part of the real Poynting vector $\text{Re}(\mathbf{\Pi})$) in the case of diagonally-polarized waves with $m_1 = -m_2 = 1$, i.e., the Stokes parameters $(\tau, \chi, \sigma) = (0, 1, 0)$ (cf. the Fedorov–Imbert transverse momentum in evanescent waves [10]). **(b)** Distribution of the $y$-component of the imaginary Poynting momentum density $\text{Im}[\Pi_y(x)]$ (13) for circularly-polarized waves with $m_1 = -m_2 = i$, i.e., the Stokes parameters $(\tau, \chi, \sigma) = (0, 0, 1)$. Despite the planar two-wave interference, both these non-canonical momenta have the *transverse y*-directed components, which are strongly *polarization-dependent*. Namely, the distributions in (a) and (b) are flipped when the polarizations are changed to the opposite: $\chi = -1$ and $\sigma = -1$. An absorbing probe particle is shown here at the $x$-position corresponding to $\Phi = \pi$, and the optical forces and torques are indicated schematically. The spin and imaginary-Poynting momenta do not exert radiation pressure in the dipole-coupling approximation, but do cause a weak polarization-dependent optical force (16) [shown in red] $\delta F_y \propto \text{Re}(\Pi_y) = \mathcal{P}_{Sy}$ and $\delta F_y \propto \text{Im}(\Pi_y)$ in the higher-order approximation (see Fig. 4 for numerical simulations). The particle also experiences the action of the in-plane radiation-pressure force (14) and torque (15) [shown in grey]; the latter

corresponds to the helicity-dependent $x$-directed spin (11) of the interfering waves with opposite circular polarizations [36].

Next, the first term in equation (12) and the second one in equation (13) [shown in orange and green frames] describe the real and imaginary parts of the *transverse $y$-directed complex Poynting momentum* (7): $\text{Re}(\Pi_y) = \mathcal{P}_{Sy}$ and $\text{Im}(\Pi_y)$, Fig. 3. In contrast to the longitudinal canonical momentum (10), both these transverse parts are strongly *polarization-dependent* ($\chi$- and $\sigma$-dependent for the chosen configuration $m_1 = -m_2$, see Supplemental Material for other cases). Importantly, the real part of the transverse Poynting vector is a pure *Belinfante's spin momentum* (8), which is analogous to the helicity-dependent transverse momentum in evanescent waves (first found by Fedorov and Imbert [14,15] and explained only very recently [10]). It emerges because of the spatial $x$-inhomogeneity of the usual longitudinal $z$-directed spin $S_z$, Eq. (11), which is in turn produced by the field rotation in the $(x,y)$ plane, Eq. (6). Both parts of $\Pi_y$ do not transport energy, do not exert the usual optical pressure, and have sinusoidal distributions across the interference fringes, Fig. 3. Nonetheless, below we show that they do reveal themselves in light-matter interactions, and, hence, can be detected experimentally.

## 3. Mechanical action on probe particles

We now describe manifestations of the unusual dynamical characteristics (9)–(13) of the two-wave field in light-matter interactions. For this purpose, we consider the field interaction with a small spherical probe particle. This approach is verified in numerous experimental [18,20–22] and theoretical [10,23–27] studies. The particle is absorptive and is characterized by subwavelength radius $r$ and complex electric and magnetic polarizabilities $\alpha^{e,m}$.

For typical non-magnetic materials, the electric polarizability is much higher than the magnetic one, because $\alpha^e \propto (kr)^3$ and $\alpha^m \propto (kr)^5$ for $kr \ll 1$ (see Appendix A and Supplemental Material). Therefore, for Rayleigh particles, the leading-order light-matter interaction is the *electric-dipole* coupling. It results in the following optical force **F** and torque **T** on the particle [10,23–27]:

$$\mathbf{F} = g^{-1}\left[\frac{1}{2\omega\varepsilon}\text{Re}(\alpha^e)\nabla W^e + \mu\,\text{Im}(\alpha^e)\mathbf{P}^e\right] \equiv \mathbf{F}_{\text{grad}} + \mathbf{F}_{\text{press}}, \quad (14)$$

$$\mathbf{T} = g^{-1}\text{Im}(\alpha^e)\mathbf{S}^e, \quad (15)$$

where the force consists of the gradient and radiation-pressure contributions. The radiation-pressure force in (14) and torque (15), both proportional to $\text{Im}(\alpha^e)$, characterize the rate of the momentum and AM transfer from light to the particle via the photon-absorption mechanism.

Being proportional to the electric parts of the canonical momentum density (5) $\mathbf{P}^e$ and spin AM density (6) $\mathbf{S}^e$, the radiation-pressure force and torque on the particle naturally *measure* these local characteristics of the field [7,10,21,22,27], including the transverse spin $S_y^e(x)$. In addition to the asymptotic analytical results (15), Figure 4a shows exact numerical calculations of the torque on a gold Mie particle, with $0 < kr < 4$, suspended in water (see Appendix A) and interacting with the two-wave interference field with linear polarizations $m = 0$ and $m = \infty$ (Stokes parameter $\tau = \pm 1$) at the $x$-position corresponding to $\Phi = \pi/2$. One can clearly see the *transverse torque* $T_y$ about the $y$-axis revealing the transverse spin AM density (11) $S_y^e(x)$, see Fig. 2, which will change the sign for the $x$-position, corresponding to $\Phi = 3\pi/2$. The particle simultaneously experiences the action of the in-plane radiation-pressure and gradient forces in this location, but these could be balanced in experiments trapping the particle at a desired $(x,z)$ point.

Remarkably, in the above dipole-coupling approximation, the transverse components of the complex Poynting momentum, equations (12) and (13), have no effect on the particle. However, they do appear in the higher-order interaction involving cross electric-magnetic dipole-dipole terms proportional to $\alpha^e \alpha^{m*} \propto (kr)^8$ [10]. Such weak interaction is negligible for Rayleigh particles, but it becomes noticeable for larger Mie particles with $kr \sim 1$. This results in the weak force correction [10,24,25]:

$$\delta \mathbf{F} = g^{-1} \frac{k^3}{3} \left[ -\text{Re}(\alpha^e \alpha^{m*}) \text{Re}(\mathbf{\Pi}) + \text{Im}(\alpha^e \alpha^{m*}) \text{Im}(\mathbf{\Pi}) \right]. \tag{16}$$

The in-plane $(x,z)$ components of this force are negligible compared to the radiation-pressure and gradient forces (14), but the transverse component $\delta F_y$ is the *only* force in the $y$-direction. It has two contributions, proportional to the transverse Belinfante's spin momentum (12), $\text{Re}(\Pi_y) = \mathcal{P}_{Sy}$, and to the imaginary Poynting momentum (13), $\text{Im}(\Pi_y)$. Therefore, both of these transverse-force components are *strongly polarization-dependent*, and are proportional to the Stokes parameters $\chi$ and $\sigma$, respectively.

Figures 4b and 4c depict the results of the exact numerical calculations of the forces exerted on the same golden Mie particle (see Appendix A) in the two-wave interference fields with polarizations $m = \pm 1$ and $m = \pm i$ (Stokes parameters $\chi = \pm 1$ an $\sigma = \pm 1$) at the $x$-position, corresponding to $\Phi = \pi$. Alongside the strong polarization-independent radiation-pressure force $F_z$, one can see extraordinary transverse forces $F_y$ changing their signs upon the flip of the Stokes parameters $\chi$ and $\sigma$ (these forces also have the opposite signs at the $x$-position, corresponding to $\Phi = 0$). Importantly, for Mie particles with $kr \sim 1$, the weak force $F_y$ is only one order of magnitude below the usual radiation-pressure force $F_z$, and, therefore, it is clearly detectable in standard optical-manipulation experiments. This proves the observability of the transverse spin momentum (12) and imaginary Poynting momentum (13), Fig. 3.

It is worth remarking that the directions of the forces depend on the parameters of the particle, and Figures 4b and 4c show the transverse forces $F_y$ corresponding to the negative factors $\text{Re}(\alpha^e \alpha^{m*}) < 0$ and $\text{Im}(\alpha^e \alpha^{m*}) < 0$ for the chosen gold particle (see Appendix A). Note also that there is no gradient force at $\Phi = 0, \pi$, which offers a natural trapping of the particle in the $x$-positions with maximum transverse forces. (The stable or unstable character of these positions depends on the sign of the gradient forces in their vicinity, which in turn is determined by the parameters of the particle [18].)

Strikingly, the field characteristics (9)–(13), dipole interactions (14) and (15), and the weak force correction (16) are not merely leading-order terms in a series of multiple light-matter interaction orders. Being asymptotic with respect to the particle size $kr$ at $kr \ll 1$, equations (9)–(16) *precisely* keep the dependencies of optical forces and torques on wave polarizations $(\tau, \chi, \sigma)$ and phases $\Phi$ even for larger Mie particles with $kr > 1$. This can be seen in detailed numerical analysis given in Supplemental Material. Thus, the above description is indeed fundamental and complete.

We also note from Eqs. (11)–(13) that all the transverse spin and momentum phenomena discussed here depend on the angle $\gamma$ as $\propto \sin 2\gamma$, which enters as an overall scaling factor for these phenomena. For the numerical simulations, we chose the reasonably small angle $\gamma = 0.1$ (paraxial propagation), which provides the period of the fringes, $\pi/(k \sin \gamma)$, of about 5 wavelengths. This is sufficient for placing a wavelength-order probe in the required $x$-position between the fringes.

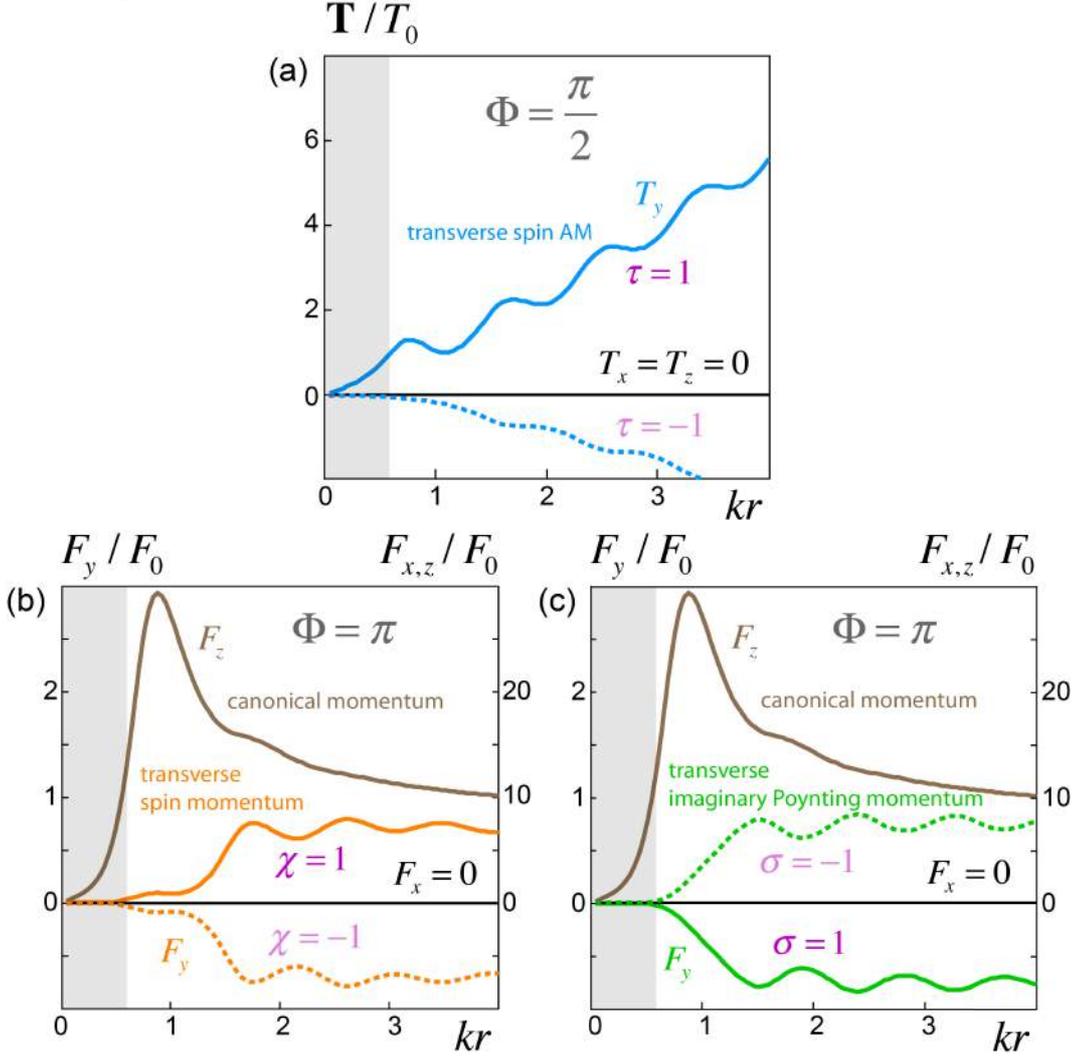

**Figure 4. Optical forces and torques on a gold Mie particle in the two-wave interference field.** Exact numerical calculations of the forces and torques exerted by the two-wave interference field (2) with $\gamma = 0.1$ and different polarizations on a gold Mie particle suspended in water (see Appendix A). The forces and torques (normalized by the factors $F_0 = gr^2 A^2$ and $T_0 = F_0/k$) are plotted as functions of the dimensionless radius of the particle, $kr$. The light-grey areas schematically indicate the Rayleigh dipole-approximation range $kr \ll 1$. **(a)** Optical torques for the simplest case of linear polarizations $m = 0$ and $m = \infty$, i.e., the Stokes parameters $\tau = \pm 1$, and particle's $x$-position corresponding to $\Phi = \pi/2$. The strong transverse torque clearly indicates the presence of a transverse helicity-independent spin AM: $T_y \propto S_y^e = S_y$ for the in-plane polarization $\tau = 1$, see equations (11) and (15) and Fig. 2. The same torque becomes weak and negative for the $\tau = -1$ polarization, because in this case the transverse spin has magnetic origin, $S_y = S_y^m$, and it is weakly coupled to the non-magnetic gold particle. **(b)** and **(c):** Optical forces for the diagonal and circular wave polarizations $m = \pm 1$ and $m = \pm i$, i.e., the Stokes parameters $\chi = \pm 1$ and $\sigma = \pm 1$, and particle's $x$-position corresponding to $\Phi = \pi$. The strongest force is the longitudinal radiation-pressure force proportional to the $z$-directed canonical momentum of light: $F_z \propto P_z^e$, equations (10) and (14). The gradient force is absent because the particle is located in the maximum of the electric-field intensity: $F_x \propto \nabla_x W^e = 0$. Importantly, a weak polarization-dependent transverse force $F_y$ appears beyond the Rayleigh dipole approximation, i.e., at $kr \sim 1$. This force (16) indicates the presence of the transverse $\chi$-dependent Belinfante's spin momentum

$\delta F_y \propto \text{Re}(\Pi_y) = P_{Sy}$ and $\sigma$-dependent imaginary Poynting momentum $\delta F_y \propto \text{Im}(\Pi_y)$, equations (12) and (13), and Fig. 3.

## 4. Conclusions

To summarize, we have shown that one of the simplest optical systems – two interfering plane waves – still provides surprises, exhibiting rather rich and unexpected local dynamical properties. Despite the seemingly planar character of the system, we have found that the two-wave field carries a non-zero transverse (out-of-plane) spin and momentum densities. These quantities appear because the planar wave vectors determine only the extrinsic degrees of freedom of the wave field, while the polarization degrees of freedom and their associated properties remain truly three-dimensional.

Remarkably, the transverse spin AM is independent of the wave helicity and appears even for linearly-polarized waves. On the contrary, the transverse component of the complex Poynting momentum is strongly polarization-dependent, the real part of this transverse momentum being Belinfante's spin momentum. The transverse Poynting momentum does not exert radiation pressure in dipole-coupling interactions, but it appears in the higher-order interactions with larger Mie particles and can be detected. We performed exact numerical calculations of the forces and torques exerted on a particle in the interference field and proved the straightforward observability of the above extraordinary dynamical features.

The fact that the transverse spin and momentum can be obtained from simple planar propagating fields is very important for experiments, as the latter are much easier to generate and design than evanescent fields considered in previous works [9–15]. Our findings offer a new vision for the fundamental properties of propagating optical fields and pave the way for nontrivial optical manipulations of small particles.

*Note added in proof*: After this work was accepted, two experimental papers [37,38] appeared, which confirm our theoretical findings. First, a weak helicity-dependent transverse force was detected in [37] for a Mie particle in interference of the orthogonal linear- and circularly-polarized plane waves. Although this force was interpreted there using "Aharonov-Bohm effect" arguments, it can be clearly explained as the force (16) proportional to the transverse imaginary part of the Poynting vector. The second term in Eq. (S10) in the Supplemental Material, with $2\gamma = \pi/2$, $m_1 = i\sigma$, and $m_2 = 0$, describes this transverse Poynting-vector component. Second, the transverse helicity-independent spin AM density was measured in [38] in a focused radially-polarized Gaussian beam. Since the beam is produced by interference of multiple plane waves, our two-wave system can be considered as a toy model for the interference phenomena in the beam [39].

## Acknowledgements

This work was partially supported by the RIKEN iTHES Project, MURI Center for Dynamic Magneto-Optics, JSPS-RFBR contract no. 12-02-92100 and a Grant-in-Aid for Scientific Research (S).

## Appendix A: Calculations of optical forces and torques on Mie particles

For numerical simulations, we consider the two-wave interference field (2) with the vacuum wavelength $\lambda_0 = 2\pi c / \omega = 650$ μm and angle $\gamma = 0.1$ interacting with a spherical gold particle of radius $r$ suspended in water. Accordingly, the medium (water) is characterized by $\varepsilon = 1.77$ and $\mu = 1$, whereas the particle electric and magnetic constants are $\varepsilon_p = -12.2 + 3.0i$ and $\mu_p = 1$.

To estimate the magnitudes of the optical forces and torques in the small-particle approximation $kr \ll 1$, equations (14)–(16), one can use equations (S26) of the Supplemental Material, which describe the effective electric and magnetic polarizabilities of a non-magnetic particle in the leading order in $kr$. With the above parameters, this yields

$$\alpha^e \simeq \left(1.28 \cdot 10^{-3} + 1.56 \cdot 10^{-4} i\right)(kr)^3 \text{ μm}^3,$$

$$\alpha^m \simeq \left(-1.24 \cdot 10^{-4} + 2.66 \cdot 10^{-5} i\right)(kr)^5 \text{ μm}^3,$$

$$\alpha^e \alpha^{m*} \simeq \left(-1.55 \cdot 10^{-7} - 5.37 \cdot 10^{-8} i\right)(kr)^8 \text{ μm}^6. \tag{A1}$$

For exact calculations with a particle of arbitrary radius $r$, we used the standard Mie theory [40], generalized for the case of two incident plane waves. Namely, using the Mie solution for a single plane wave, we determine the scattered electromagnetic fields $(\mathbf{E}_1^s, \mathbf{H}_1^s)$ and $(\mathbf{E}_2^s, \mathbf{H}_2^s)$ for each of the incident waves, $(\mathbf{E}_1, \mathbf{H}_1)$ and $(\mathbf{E}_2, \mathbf{H}_2)$. Hence, the total field, perturbed by the interaction with the particle, is

$$\mathbf{E}^{tot} = \mathbf{E}_1 + \mathbf{E}_2 + \mathbf{E}_1^s + \mathbf{E}_2^s,$$

$$\mathbf{H}^{tot} = \mathbf{H}_1 + \mathbf{H}_2 + \mathbf{H}_1^s + \mathbf{H}_2^s. \tag{A2}$$

Once the total field is known, its mechanical action on the particle is calculated via the standard procedures using the Maxwell stress tensor [2] $\hat{Q} = \{Q_{ij}\}$, $i,j = x,y,z$:

$$Q_{ij} = g\omega \operatorname{Re}\left[\varepsilon E_i^{tot*} E_j^{tot} + \mu H_i^{tot*} H_j^{tot} - \frac{1}{2}\delta_{ij}\left(\varepsilon |\mathbf{E}^{tot}|^2 + \mu |\mathbf{H}^{tot}|^2\right)\right], \tag{A3}$$

and the corresponding AM flux tensor $\hat{M} = \{M_{ij}\}$, $M_{ij} = e_{jkl} x_k Q_{li}$. Here $\delta_{ij}$ is the Kronecker delta, $e_{jkl}$ is the Levi–Civita symbol, and $\{x_i\} = \{x,y,z\}$. Integrating the stress tensor and the AM flux tensor components over any surface $\Sigma$ enclosing the particle (e.g., a sphere with radius $R > r$), we obtain the optical force and torque exerted on the particle:

$$\mathbf{F} = \oint_\Sigma \hat{Q} \mathbf{n} \, d\Sigma = R^2 \int_\Omega \hat{Q} \mathbf{n} \, d\Omega,$$

$$\mathbf{T} = \oint_\Sigma \hat{M} \mathbf{n} \, d\Sigma = R^2 \int_\Omega \hat{M} \mathbf{n} \, d\Omega. \tag{A4}$$

Here $d\Omega = \sin\theta \, d\theta \, d\phi$ is the elementary solid angle, and $\mathbf{n} = (\sin\theta\cos\phi, \sin\theta\sin\phi, \cos\theta)^T$ is the unit vector of the outer normal to the surface of the sphere. Finally, the forces and torques calculated using the above method are normalized by the factors $F_0 = gr^2 A^2$ and $T_0 = F_0/k$, and are plotted in Fig. 4 as well as in Figs. S2 and S3 in Supplemental Material. Since $\alpha^e \propto (kr)^3$ in the Rayleigh limit $kr \ll 1$, such normalization implies *linear* growth with $kr$, at $kr \ll 1$, for the electric-dipole quantities (14) and (15).

# Supplemental Material

## 1. Wave fields.

The complex electric and magnetic field amplitudes of the two waves propagating with the wave vectors $\mathbf{k}_{1,2} = k(\cos\gamma\,\bar{\mathbf{z}} \pm \sin\gamma\,\bar{\mathbf{x}})$ (see Fig. 1 of the main text), are as follows:

$$\mathbf{E}_{1,2} = \frac{A}{\sqrt{1+|m_{1,2}|^2}}\left(\cos\gamma\,\bar{\mathbf{x}} + m_{1,2}\,\bar{\mathbf{y}} \mp \sin\gamma\,\bar{\mathbf{z}}\right)e^{i\Phi_{1,2}}, \qquad (S1a)$$

$$\mathbf{H}_{1,2} = \sqrt{\frac{\varepsilon}{\mu}}\frac{A}{\sqrt{1+|m_{1,2}|^2}}\left(-m_{1,2}\cos\gamma\,\bar{\mathbf{x}} + \bar{\mathbf{y}} \pm m_{1,2}\sin\gamma\,\bar{\mathbf{z}}\right)e^{i\Phi_{1,2}}. \qquad (S1b)$$

Here $\Phi_{1,2} = k(z\cos\gamma \pm x\sin\gamma)$ are the wave phases, the waves have equal amplitudes $A$, and $m_{1,2}$ are the complex parameters describing the wave polarizations [10,28]. The normalized Stokes parameters $(\tau,\chi,\sigma)$, $\tau^2 + \chi^2 + \sigma^2 = 1$, representing the polarization on the Poincaré sphere, are:

$$\tau_{1,2} = \frac{1-|m_{1,2}|^2}{1+|m_{1,2}|^2}, \qquad \chi_{1,2} = \frac{2\,\mathrm{Re}\,m_{1,2}}{1+|m_{1,2}|^2}, \qquad \sigma_{1,2} = \frac{2\,\mathrm{Im}\,m_{1,2}}{1+|m_{1,2}|^2}. \qquad (S2)$$

These parameters represent the degrees of the vertical/horizontal, diagonal $45°/-45°$, and right-hand/left-hand circular polarizations, respectively, so that $\sigma_{1,2}$ are the helicities of the two waves. In the main text we consider two waves with the "mirror-symmetric" polarization states:

$$m_1 = -m_2 \equiv m, \quad (\tau_1,\chi_1,\sigma_1) = (\tau_2,-\chi_2,-\sigma_2) \equiv (\tau,\chi,\sigma), \qquad (S3)$$

and the reason for such choice is explained below.

The resulting wave interference field is

$$\mathbf{E} = \mathbf{E}_1 + \mathbf{E}_2, \qquad \mathbf{H} = \mathbf{H}_1 + \mathbf{H}_2. \qquad (S4)$$

This field propagates along the $z$-axis and is inhomogeneous along the $x$-coordinate, Fig. 1. The transverse inhomogeneity is described by the relative phase $\Phi = \Phi_1 - \Phi_2 = 2kx\sin\gamma$, and the problem is $2\pi$-periodic in $\Phi$. To illustrate the evolution of the electric and magnetic fields during the wave propagation, it is natural to plot the field distributions along the $z$-axis for different values of $x$ or $\Phi(x)$. The case of linear in-plane polarization $m = 0$ (i.e., $\tau = 1$) is shown in Fig. 2 of the main text. In Figure S1, we depict such instantaneous $t = 0$ distributions of the real electric and magnetic fields $\mathcal{E}(\mathbf{r},t) = \mathrm{Re}\left[\mathbf{E}(\mathbf{r})e^{-i\omega t}\right]$ and $\mathcal{H}(\mathbf{r},t) = \mathrm{Re}\left[\mathbf{H}(\mathbf{r})e^{-i\omega t}\right]$ for three basic polarization states: $m = 0$, $m = 1$ (diagonal polarizations $\chi = 1$ corresponding to Fig. 3a), and $m = i$ (circular polarizations $\sigma = 1$ corresponding to Figs. 1 and 3b).

Various nontrivial rotations of the electric and magnetic fields in the propagation of this fields along the $z$-axis perfectly explain all components of the electric and magnetic spin AM densities in equation (11) of the main text. These are:
(i) the longitudinal spin AM densities $S_z^{e,m} \propto \chi\sin\Phi$, which are maximal and minimal at the diagonal polarizations $m = \pm 1$ ($\chi = \pm 1$);

(ii) the transverse in-plane (horizontal) spin AM densities $S_x^{e,m} \propto \sigma(1 \mp \sin\Phi)$, which are maximal and minimal at the circular polarizations $m = \pm i$ ($\sigma = \pm 1$);

(iii) the transverse out-of-plane spin AM densities $S_y^{e,m} \propto (\tau \pm 1)\sin\Phi$, which are maximal and minimal at the TM and TE polarizations $m = 0$ and $m = \infty$ ($\tau = \pm 1$), see Figs. 2 and 3a.

Note that the spin component (ii) is the only spin AM that does not vanish after averaging over the $x$-coordinate [i.e., over $\Phi \in (0, 2\pi)$]. This is natural because the interfering waves with opposite helicities $\sigma_1 = -\sigma_2$ and opposite $x$-components of their wave vectors $k_{1x} = -k_{2x}$ have the same $x$-components of their spin AM. Of course, the spin AM is not additive in the interference (it is quadratic in fields), but this consideration makes the appearance of the $x$-directed spin AM intuitively clear. Similar in-plane transverse AM was recently considered in [36] using the tight focusing of two half-beams with opposite circular polarizations.

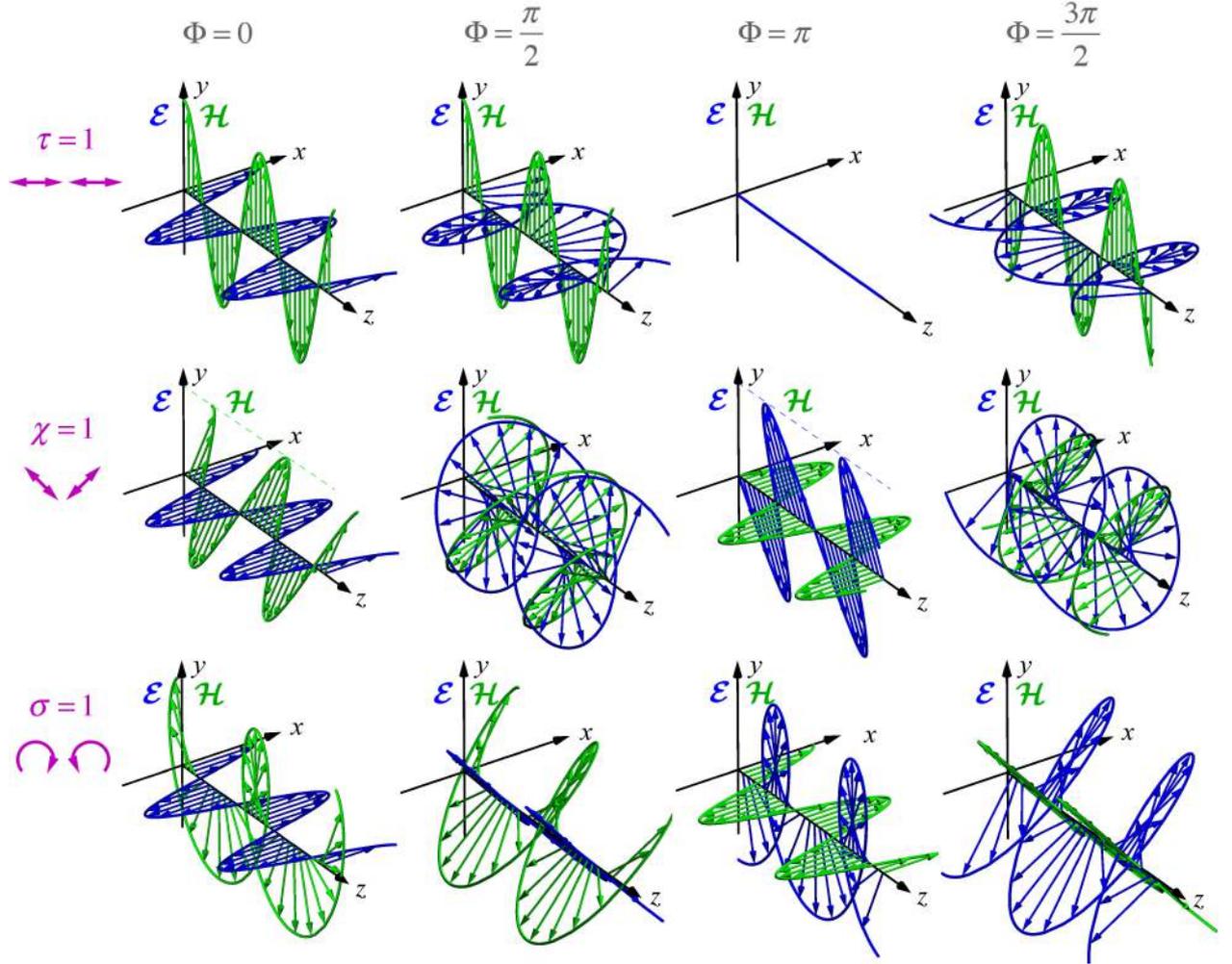

**Figure S1.** Instant distributions of the real electric and magnetic fields $\mathcal{E}(\mathbf{r},0) = \text{Re}[\mathbf{E}(\mathbf{r})]$ and $\mathcal{H}(\mathbf{r},0) = \text{Re}[\mathbf{H}(\mathbf{r})]$ in two-wave interference, equations (S1)–(S4), with different polarizations (S3). These distributions are independent of $y$ and are plotted as functions of the longitudinal $z$-coordinate at different values of $x$, indicated by the values of the phase $\Phi = 2kx\sin\gamma$. The polarizations $m = 0$, $m = 1$, and $m = i$ correspond to the in-plane linear polarization, diagonal 45° polarization, and circular polarization with the Stokes parameters $(\tau, \chi, \sigma) = (1,0,0)$, $(0,1,0)$, and $(0,0,1)$, respectively.

## 2. Dynamical characteristics of the interference field.

The main local dynamical characteristics of an electromagnetic field are the energy, momentum, and spin densities, given by equations (4)–(6), and also the Poynting and spin momentum (7) and (8). For the sake of completeness, we also include here the density of *helicity* $K$, which is an independent meaningful characteristic of optical fields (see [27,29,S1–S8] for recent studies). The helicity density of a monochromatic field is

$$K = -g \, \text{Im}\left(\mathbf{E}^* \cdot \mathbf{H}\right), \tag{S5}$$

Substituting the two-wave superposition fields (S1), (S2), and (S4) into general equations (2)–(6) and (S5) and omitting the common factor $\varepsilon g A^2$ in all dynamical characteristics, we calculate the energy, momentum, and spin distributions in the field:

$$W^e = \frac{\omega}{N}\left\{N + \text{Re}\left[\left(\cos 2\gamma + m_1 m_2^*\right)e^{i\Phi}\right]\right\}, \quad W^m = \frac{\omega}{N}\left\{N + \text{Re}\left[\left(m_1 m_2^* \cos 2\gamma + 1\right)e^{i\Phi}\right]\right\}, \tag{S6}$$

$$\mathbf{P}^e = \frac{\cos\gamma}{cn} W^e \, \overline{\mathbf{z}}, \quad \mathbf{P}^m = \frac{\cos\gamma}{cn} W^m \, \overline{\mathbf{z}}, \tag{S7}$$

$$\mathbf{S}^e = \frac{1}{2n^2 N}\left\{\sin\gamma\left[(\sigma_1-\sigma_2)N - 2\,\text{Im}\left[(m_1+m_2^*)e^{i\Phi}\right]\right]\overline{\mathbf{x}} + \boxed{2\sin 2\gamma \,\text{Im}(e^{i\Phi})\overline{\mathbf{y}}}\right.$$
$$\left. +\cos\gamma\left[(\sigma_1+\sigma_2)N + 2\,\text{Im}\left[(m_1-m_2^*)e^{i\Phi}\right]\right]\overline{\mathbf{z}}\right\}, \tag{S8a}$$

$$\mathbf{S}^m = \frac{1}{2n^2 N}\left\{\sin\gamma\left[(\sigma_1-\sigma_2)N + 2\,\text{Im}\left[(m_1+m_2^*)e^{i\Phi}\right]\right]\overline{\mathbf{x}} + \boxed{2\sin 2\gamma \,\text{Im}(m_1 m_2^* e^{i\Phi})\overline{\mathbf{y}}}\right.$$
$$\left. +\cos\gamma\left[(\sigma_1+\sigma_2)N + 2\,\text{Im}\left[(m_1-m_2^*)e^{i\Phi}\right]\right]\overline{\mathbf{z}}\right\}, \tag{S8b}$$

$$\mathcal{P}_s = \frac{k \sin 2\gamma}{n^2 N}\left\{\boxed{-\text{Re}\left[(m_1-m_2^*)e^{i\Phi}\right]\overline{\mathbf{y}}} + \sin\gamma \,\text{Re}\left[(1+m_1 m_2^*)e^{i\Phi}\right]\overline{\mathbf{z}}\right\}, \tag{S9}$$

$$\text{Im}(\mathbf{\Pi}) = -\frac{k}{n^2 N}\left\{2\sin\gamma \,\text{Im}\left[(1-m_1 m_2^*)e^{i\Phi}\right]\overline{\mathbf{x}} + \boxed{\sin 2\gamma \,\text{Im}\left[(m_1+m_2^*)e^{i\Phi}\right]\overline{\mathbf{y}}}\right\}. \tag{S10}$$

$$K = \frac{1}{nN}\left\{(\sigma_1+\sigma_2)N + 2\cos^2\gamma \,\text{Im}\left[(m_1-m_2^*)e^{i\Phi}\right]\right\}, \tag{S11}$$

where $N = \sqrt{1+|m_1|^2}\sqrt{1+|m_2|^2}$. Here the coloured frames indicate the main subjects of our study, cf. equations (11)–(13) in the main text: the transverse spin AM density in (S8), the transverse Belinfante's spin momentum density in (S9), and the transverse imaginary Poynting momentum in (S10). It is worth noticing that performing spatial averaging of equations (S6)–(S11) over the phase $\Phi \in (0, 2\pi)$, we obtain a natural picture of these quantities as for two non-interfering photons with wave vectors $\mathbf{k}_{1,2} = k(\cos\gamma \,\overline{\mathbf{z}} \pm \sin\gamma \,\overline{\mathbf{x}})$ and helicities $\sigma_{1,2}$ (ignoring the refractive index $n$ of the medium):

$$\langle W^{e,m}\rangle = \omega, \quad \langle \mathbf{P}^{e,m}\rangle = \frac{k\cos\gamma}{n^2}\,\overline{\mathbf{z}}, \tag{S12}$$

$$\langle \mathbf{S}^{e,m}\rangle = \frac{1}{2n^2}\left\{\sin\gamma(\sigma_1-\sigma_2)\overline{\mathbf{x}} + \cos\gamma(\sigma_1+\sigma_2)\overline{\mathbf{z}}\right\}, \quad \langle K\rangle = \frac{1}{n}(\sigma_1+\sigma_2), \tag{S13}$$

and $\langle \mathcal{P}_s \rangle = \langle \text{Im}(\Pi) \rangle = 0$.

In the case of the "mirror" polarizations (S3), equations (S6)–(S10) are simplified to equations (9)–(13) of the main text, whereas the helicity density (S11) yields

$$K = \frac{2}{n}\chi \cos^2\gamma \sin\Phi. \tag{S14}$$

In the case of *equal* polarizations of the two interfering waves,

$$m_1 = m_2 \equiv m, \qquad (\tau_1, \chi_1, \sigma_1) = (\tau_2, \chi_2, \sigma_2) \equiv (\tau, \chi, \sigma), \tag{S15}$$

equations (S5)–(S9) become

$$W^{\text{e,m}} = \omega\left[1 + \left(\cos^2\gamma \mp \tau \sin^2\gamma\right)\cos\Phi\right], \qquad \mathbf{P}^{\text{e,m}} = \frac{\cos\gamma}{cn}W^{\text{e,m}}\,\overline{\mathbf{z}}, \tag{S16}$$

$$\mathbf{S}^{\text{e,m}} = \frac{1}{n^2}\left[\mp\chi\sin\gamma\sin\Phi\,\overline{\mathbf{x}} + \boxed{\frac{1\pm\tau}{2}\sin 2\gamma\sin\Phi\,\overline{\mathbf{y}}} + \sigma\cos\gamma\left(1+\cos\Phi\right)\overline{\mathbf{z}}\right], \tag{S17}$$

$$\mathcal{P}_s = \frac{k\sin 2\gamma}{n^2}\left[\boxed{\sigma\sin\Phi\,\overline{\mathbf{y}}} + \sin\gamma\cos\Phi\,\overline{\mathbf{z}}\right], \tag{S18}$$

$$\text{Im}(\Pi) = -\frac{2k\sin\gamma}{n^2}\left[\tau\sin\Phi\,\overline{\mathbf{x}} + \boxed{\chi\cos\gamma\sin\Phi\,\overline{\mathbf{y}}}\right]. \tag{S19}$$

$$K = \frac{2}{n}\sigma\left(1+\cos^2\gamma\cos\Phi\right), \tag{S20}$$

where the two signs correspond to the indices "e" and "m".

One can see that in the case of equal polarizations (S13) the field has similar unusual local spin and momentum properties [marked by coloured frames here] as in equations (11)–(13) of the main text. A slight difference is that the unusual transverse momenta in (S18) and (S19) appear now with factors of $\sin\Phi$ rather than $\cos\Phi$. This means that these momenta vanish in the regions of maximal and minimal intensity of the interference field (varying as $1+\cos\Phi$). However, in optical manipulation experiments, particles are usually trapped in the regions of maximal or minimal intensity, and, therefore, the equal-polarization configuration (S13) is less favourable for such experiments than the mirror-polarization (S3), which is considered in the main text. This explains our choice of the wave polarizations in this work.

## 3. Interaction of the field with a small spherical particle.

Detailed considerations of the interaction of a monochromatic optical field with a small isotropic spherical particle can be found in [23–25] and in the Supplementary Information of [10]. Therefore, here we will only summarize the main results for the optical forces and torques acting on the particle. The particle is characterized by its small radius $r < k^{-1}$ and complex electric and magnetic dipole polarizabilities $\alpha^{\text{e,m}}$. The optical force and torque on a dipole particle can be written as

$$\mathbf{F} = g^{-1}\left[\frac{1}{2\omega\varepsilon}\text{Re}(\alpha^{\text{e}})\nabla W^{\text{e}} + \frac{1}{2\omega\mu}\text{Re}(\alpha^{\text{m}})\nabla W^{\text{m}} + \mu\,\text{Im}(\alpha^{\text{e}})\mathbf{P}^{\text{e}} + \varepsilon\,\text{Im}(\alpha^{\text{m}})\mathbf{P}^{\text{m}}\right], \tag{S21}$$

$$\mathbf{T} = g^{-1}\left[\mu\,\text{Im}(\alpha^{\text{e}})\mathbf{S}^{\text{e}} + \varepsilon\,\text{Im}(\alpha^{\text{m}})\mathbf{S}^{\text{m}}\right], \tag{S22}$$

where the electric and magnetic interactions are clearly separated, as well as the gradient and optical-pressure forces. Noteworthily, the energy gradients and the gradient parts of the dipole force (S21) can be associated with the imaginary part of the *complex* canonical momentum, which appears, e.g., in quantum weak measurements [7,10]:

$$\frac{g}{2}\Im\left[\frac{1}{\mu}\mathbf{E}^*\cdot(-i\nabla)\mathbf{E} + \frac{1}{\varepsilon}\mathbf{H}^*\cdot(-i\nabla)\mathbf{H}\right] = \frac{1}{2\omega\varepsilon\mu}\left[\nabla W^{\mathrm{e}} + \nabla W^{\mathrm{m}}\right], \tag{S23}$$

Equations (S21) and (S22) do not take into account the coupling between the electric and magnetic dipoles. The corresponding higher-order electric-magnetic correction to the force is [10,24,25]

$$\delta\mathbf{F} = g^{-1}\frac{k^3}{3}\left[-\operatorname{Re}(\alpha^{\mathrm{e}}\alpha^{\mathrm{m}*})\operatorname{Re}(\mathbf{\Pi}) + \operatorname{Im}(\alpha^{\mathrm{e}}\alpha^{\mathrm{m}*})\operatorname{Im}(\mathbf{\Pi})\right], \tag{S24}$$

which is equation (16) of the main text.

The electric and magnetic polarizabilities of a small spherical particle can be expressed via its radius and electromagnetic characteristics, i.e., permittivity $\varepsilon_{\mathrm{p}}$ and permeability $\mu_{\mathrm{p}}$. In the leading orders in $kr$, these expressions, obtained from the Mie scattering coefficients, are [24,25]:

$$\alpha^{\mathrm{e}} = \frac{\varepsilon}{k^3}\left\{\frac{\varepsilon_{\mathrm{p}}-\varepsilon}{\varepsilon_{\mathrm{p}}+2\varepsilon}(kr)^3 + \frac{3}{10}\frac{\varepsilon_{\mathrm{p}}^2 + \varepsilon_{\mathrm{p}}\varepsilon\left[(\varepsilon_{\mathrm{p}}\mu_{\mathrm{p}}/\varepsilon\mu)-6\right]+4\varepsilon^2}{(\varepsilon_{\mathrm{p}}+2\varepsilon)^2}(kr)^5\right\}, \tag{S25a}$$

$$\alpha^{\mathrm{m}} = \frac{\mu}{k^3}\left\{\frac{\mu_{\mathrm{p}}-\mu}{\mu_{\mathrm{p}}+2\mu}(kr)^3 + \frac{3}{10}\frac{\mu_{\mathrm{p}}^2 + \mu_{\mathrm{p}}\mu\left[(\varepsilon_{\mathrm{p}}\mu_{\mathrm{p}}/\varepsilon\mu)-6\right]+4\mu^2}{(\mu_{\mathrm{p}}+2\mu)^2}(kr)^5\right\}. \tag{S25b}$$

Usually both the particle and the surrounding medium are non-magnetic: $\mu = \mu_{\mathrm{p}} = 1$. This results in the following leading-order polarizabilities:

$$\alpha^{\mathrm{e}} \simeq \frac{1}{k^3}\frac{\varepsilon(\varepsilon_{\mathrm{p}}-\varepsilon)}{\varepsilon_{\mathrm{p}}+2\varepsilon}(kr)^3, \quad \alpha^{\mathrm{m}} = \frac{1}{k^3}\frac{(\varepsilon_{\mathrm{p}}-\varepsilon)}{30\varepsilon}(kr)^5. \tag{S26}$$

Hence $|\alpha^{\mathrm{m}}| \ll |\alpha^{\mathrm{e}}|$, and in most cases one can consider only the electric parts of the forces (S21) and torques (S22), which yields the equations (14) and (15) of the main text. At the same time, the electric-magnetic correction (S24) or (16) can be essential, when it has components along the directions where the standard forces (S22) vanish. This is precisely the case with extraordinary transverse forces proportional to Belinfante's spin momentum and the imaginary Poynting vector, as in this work and in [10] for evanescent waves. For non-magnetic polarizabilities (S26), the coefficients in the force (S24) or (16) take the form

$$\operatorname{Re}(\alpha^{\mathrm{e}}\alpha^{\mathrm{m}*}) \simeq \frac{1}{30k^6}\left|\frac{\varepsilon_{\mathrm{p}}-\varepsilon}{\varepsilon_{\mathrm{p}}+2\varepsilon}\right|^2\left[\operatorname{Re}(\varepsilon_{\mathrm{p}})+2\right](kr)^8,$$

$$\operatorname{Im}(\alpha^{\mathrm{e}}\alpha^{\mathrm{m}*}) \simeq -\frac{1}{30k^6}\left|\frac{\varepsilon_{\mathrm{p}}-\varepsilon}{\varepsilon_{\mathrm{p}}+2\varepsilon}\right|^2 \operatorname{Im}(\varepsilon_{\mathrm{p}})(kr)^8. \tag{S27}$$

Note also that the imaginary parts of the polarizabilities (S25) and (S26) vanish, together with the radiation-pressure forces in (S21) and torques (S22), in the case of non-absorbing particles with real $\varepsilon_p$ and $\mu_p$. Nonetheless, a weak optical pressure is still exerted on such non-absorbing particle (but no torque), and it is described by the following radiation-friction correction in the forces (S21): $\alpha^e \to \alpha^e + i\frac{2k^3}{3\varepsilon}|\alpha^e|^2$ and $\alpha^m \to \alpha^m + i\frac{2k^3}{3\varepsilon}|\alpha^m|^2$ [S9,S10].

The above forces and torques show only the leading-order terms in $kr$, and they are clearly connected to the fundamental local characteristics of the field: electric energy $W^e$, canonical momentum $\mathbf{P}^e$, spin AM density $\mathbf{S}^e$, and complex Poynting vector $\mathbf{\Pi}$. Rigorously speaking, the above analytical expressions for the forces and torques are applicable only to small particles with $kr \ll 1$. However, the weakness of the force (S24) or (16), proportional to $(kr)^8$, implies that it becomes noticeable only for larger Mie particles with $kr \sim 1$. For such particles, optical forces and torques can be calculated only numerically using the exact Mie theory (see Appendix). Remarkably, even for larger Mie particles with $kr > 1$, the exact optical forces and torques precisely keep all the characteristic features (dependences on polarizations and coordinates) as determined by the field characteristics energy $W^e$, $\mathbf{P}^e$, $\mathbf{S}^e$, and $\mathbf{\Pi}$. This is because the higher-order corrections to the forces mostly change the equations for the polarizability coefficients, but do not affect the proportionality to the corresponding field properties.

The most interesting cases of optical forces and torques, which reveal the presence of the extraordinary helicity-dependent transverse spin $S_y^e \propto (\tau+1)\sin\Phi$, polarization dependent transverse spin momentum $\mathcal{P}_{Sy} = \mathrm{Re}(\Pi_y) \propto \chi\cos\Phi$, and imaginary Poynting momentum $\mathrm{Im}(\Pi_y) \propto \sigma\cos\Phi$, are shown in Figure 4 of the main text. In Figures S2 and S3 we show a complete set of numerically-calculated optical forces and torques exerted on a gold Mie particle with $0 < kr < 4$, for all basic wave polarizations $\tau = \pm 1$, $\chi = \pm 1$, and $\sigma = \pm 1$, and for the particle $x$-positions corresponding to $\Phi = \pi/2$ and $\Phi = \pi$. The comparison of the exact forces and torques in Figs. S2 and S3 with the corresponding dynamical characteristics of the field is given in Table S1. Remarkably, one can trace precisely the same $(\tau, \chi, \sigma)$- and $\Phi$-dependences of all forces and torques as described by equations (9)–(13) and small-particle approximate expressions (14)–(16). This proves the physical meaningfulness, completeness, and separate observability of all dynamical properties of the interference field, which we describe in this work.

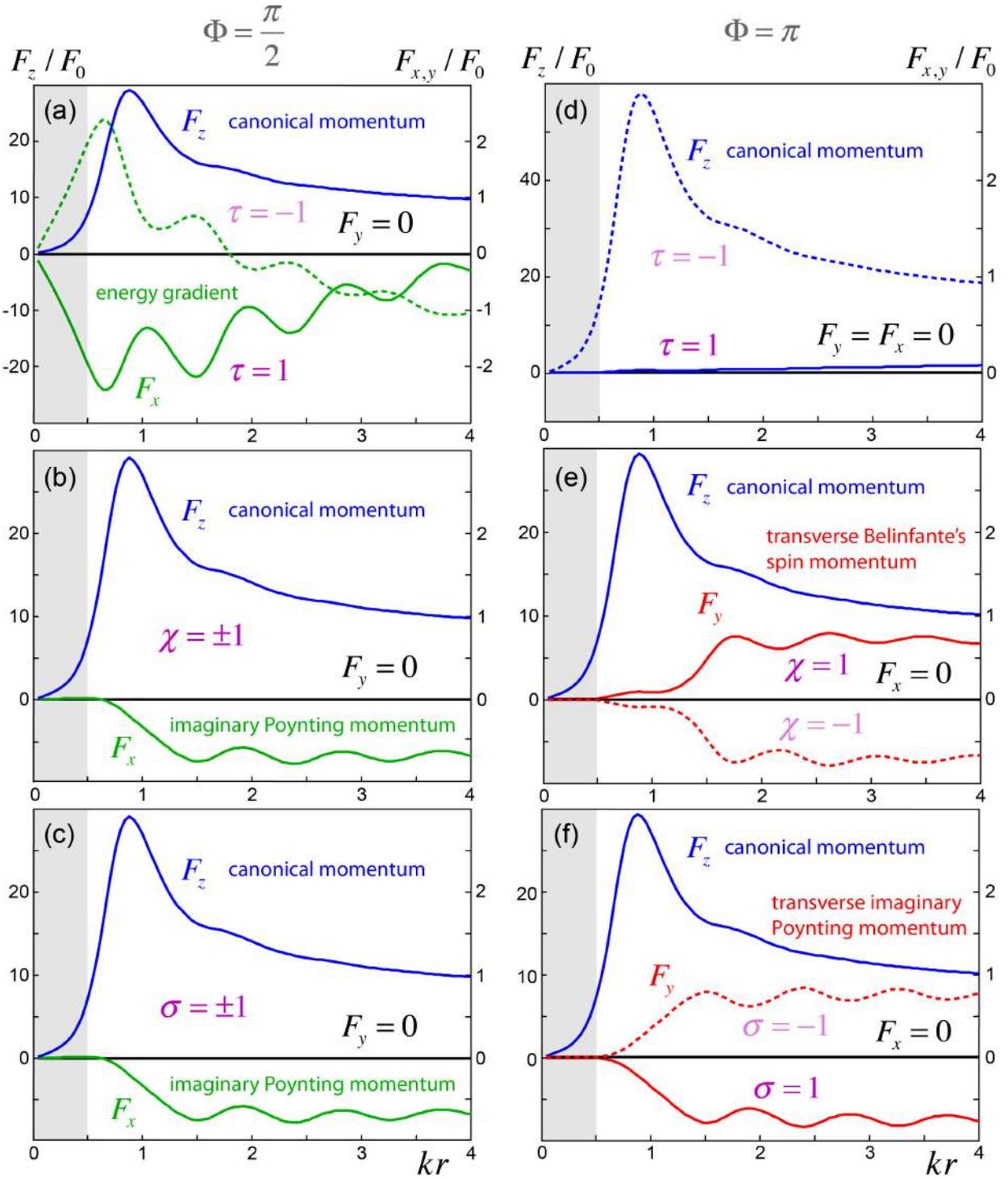

**Figure S2.** Numerically calculated force **F** exerted by a two-wave interference field (S1)–(S4) with $\gamma = 0.1$ on a gold Mie particle of radius $r$ suspended in water (see Appendix A). The cases with all basic polarizations, vertical/horizontal ($\tau = \pm 1$), diagonal 45°/-45° ($\chi = \pm 1$), and right-/left-hand circular ($\sigma = \pm 1$) are shown for two particle $x$-positions corresponding to $\Phi = \pi/2$ (on the slope of the interference fringe) and $\Phi = \pi$ (at the interference maximum). The $x$, $y$, and $z$ components of the force are shown here in green, red, and blue, whereas the dashed curves correspond to the negative values of the corresponding Stokes parameters. The light-grey areas schematically indicate the dipole-approximation range $kr \ll 1$, where the weak force (16) is negligible. The dependences of the force components on the phase $\Phi$ and wave polarizations $(\tau, \chi, \sigma)$ clearly show a direct correspondence to the energy-momentum quantities (9), (10), (12), and (13) of the interference field, as shown in the approximate formulas (14) and (16); see details in Table S1.

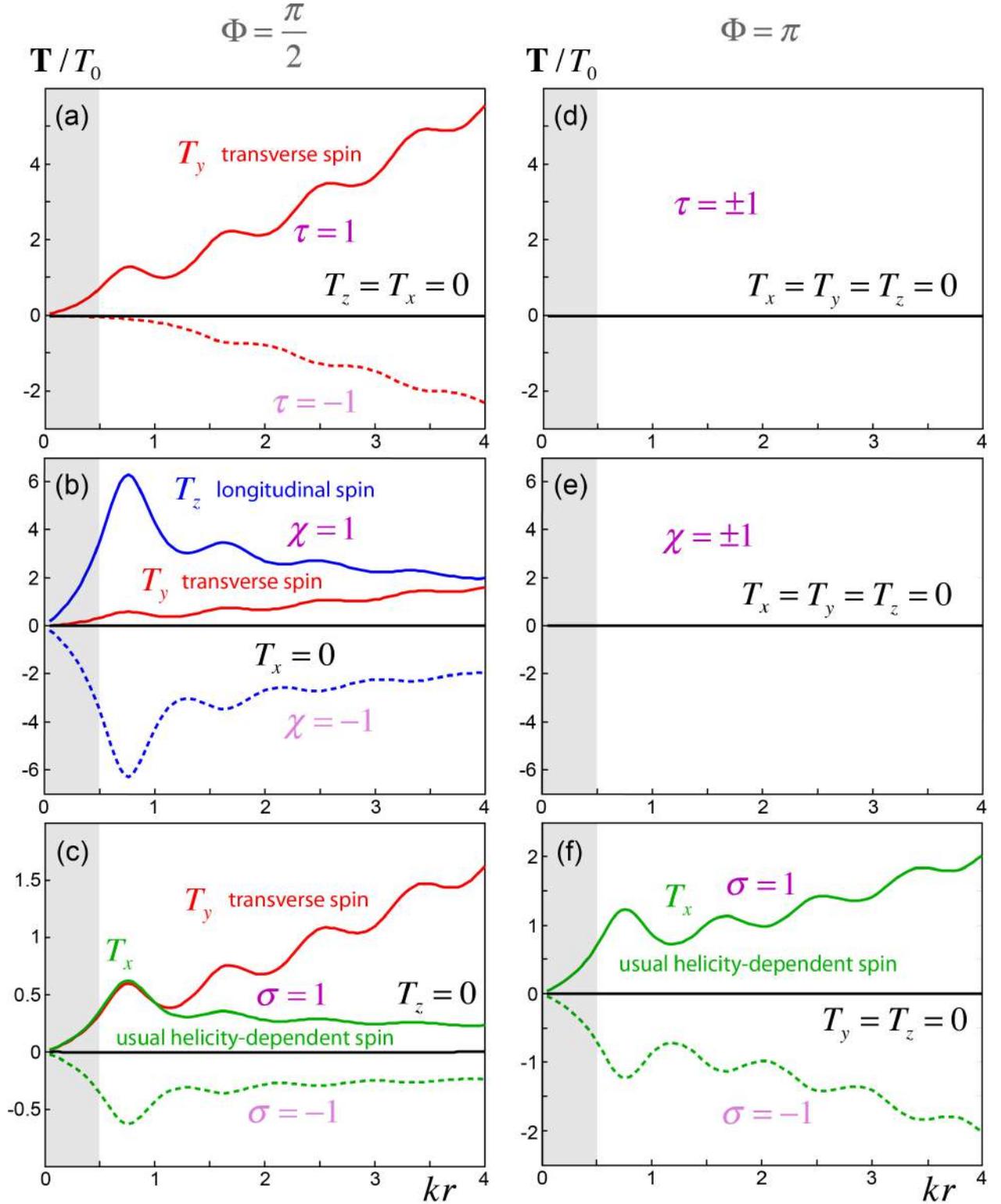

**Figure S3.** Same as in Fig. S2 but for optical torque **T** exerted on a gold Mie particle (see Appendix A). The dependence of the torque components on the $x$-dependent phase $\Phi$ and wave polarizations $(\tau, \chi, \sigma)$ clearly shows the direct correspondence to the spin AM density (11) in the interference field, as indicated in dipole-coupling approximation (15); see details in Table S1.

| **Field characteristics** | **Action on a probe particle** |
|---|---|
| Longitudinal electric canonical momentum $$P_z^e \propto 1 + \tau \cos \Phi$$ | Longitudinal $\tau$-dependent electric-dipole radiation-pressure force, Fig. S2 $$F_z \propto P_z^e$$ |
| Gradient of the electric energy density $$\nabla_x W^e \propto \tau \sin \Phi$$ | $\tau$-dependent electric-dipole gradient force, Fig. S2(a) $$F_x \propto \nabla_x W^e$$ |
| Transverse polarization-dependent Belinfante's spin momentum $$\mathcal{P}_{Sy} = \mathrm{Re}(\Pi_y) \propto \chi \cos \Phi$$ | Transverse $\chi$-dependent weak dipole-dipole force (16), Fig. S2(e) $$F_y \propto \mathcal{P}_{Sy} = \mathrm{Re}(\Pi_y)$$ |
| Transverse polarization-dependent imaginary Poynting momentum $$\mathrm{Im}(\Pi_y) \propto \sigma \cos \Phi$$ | Transverse $\sigma$-dependent weak dipole-dipole force (16), Fig. S2(f) $$F_y \propto \mathrm{Im}(\Pi_y)$$ |
| Horizontal imaginary Poynting momentum $$\mathrm{Im}(\Pi_x) \propto \sin \Phi$$ | Horizontal polarization-independent weak dipole-dipole force (16), Fig. S2(a)–(c) $$F_x \propto \mathrm{Im}(\Pi_x)$$ |
| Horizontal helicity-dependent electric spin [35] $$S_x^e \propto \sigma (1 - \cos \Phi)$$ | Horizontal $\sigma$-dependent electric-dipole torque, Fig. S3(c) and (f) $$T_x \propto S_x^e$$ |
| Transverse helicity-independent electric spin $$S_y^e \propto (\tau + 1) \sin \Phi$$ | Transverse $\tau$-dependent electric-dipole torque, Fig. S3(a)–(c) $$T_y \propto S_y^e$$ |
| Longitudinal electric spin at diagonal polarizations $$S_z^e \propto \chi \sin \Phi$$ | Longitudinal $\chi$-dependent electric-dipole torque, Fig. S3(b) $$T_z \propto S_z^e$$ |

**Table S1.** Complete set of observable momentum and spin quantities (9)–(13) which appear in two-wave interference and manifest themselves in the corresponding optical forces and torques, approximated by equations (14)–(16) for small particles. The dependences of the dynamical quantities on the wave polarization $(\tau, \chi, \sigma)$ and relative $x$-dependent phase $\Phi$ are given for the paraxial $\gamma \ll 1$ case. Precisely the same dependences can be seen in the sets of exact numerically-calculated forces and torques exerted on Mie particles, which are plotted in Figs. S2 and S3.